\documentclass[fleqn,twoside]{article}
\usepackage{espcrc2}

\usepackage{graphicx}
\usepackage[figuresright]{rotating}

\newcommand{\AmS}{{\protect\the\textfont2
  A\kern-.1667em\lower.5ex\hbox{M}\kern-.125emS}}

\newcommand{\lsim}{
\mathrel{\hbox{\rlap{\hbox{\lower4pt\hbox{$\sim$}}}\hbox{$<$}}}}

\newcommand{\gsim}{
\mathrel{\hbox{\rlap{\hbox{\lower4pt\hbox{$\sim$}}}\hbox{$>$}}}}
  
  \def\D0{D\O }

\title{B Physics in the LHC Era: Selected Topics}

\author{R. Fleischer\address[MCSD]{Nikhef, Science Park 105, 
NL-1098 XG Amsterdam, The Netherlands}%
               }
       
\begin{document}


\thispagestyle{empty}

\begin{flushright}
Nikhef-2010-034\\
\end{flushright}

\vspace{3.0truecm}
\begin{center}
\boldmath
\large\bf B Physics in the LHC Era: Selected Topics
\unboldmath
\end{center}

\vspace{0.9truecm}
\begin{center}
Robert Fleischer\\[0.1cm]
{\sl Nikhef, Science Park 105, 
NL-1098 XG Amsterdam, The Netherlands}
\end{center}

\vspace{0.9truecm}

\begin{center}
{\bf Abstract}
\end{center}

{\small
\vspace{0.2cm}\noindent
We have just entered a new round in the testing of the flavour sector of the Standard Model 
through high-precision measurements of $B$-meson decays. A particularly exciting aspect
is the exploration of the $B_s$-meson system at LHCb. We focus on two particularly 
promising probes of new physics which may give us first solid evidence for New Physics
at the LHC: the strongly suppressed rare decay $B^0_s\to\mu^+\mu^-$ and CP-violating 
effects in the $B^0_s\to J/\psi \phi$ channel. We discuss recent theoretical developments
related to these measurements and shall also sketch other highlights of the $B$-physics
programme in the LHC era. 
}

\vspace{0.9truecm}

\begin{center}
{\sl Invited talk at the 3rd Workshop on Theory, Phenomenology and Experiments in 
Heavy Flavour Physics\\
Capri, Italy, 5--7 July 2010\\
To appear in the Proceedings (Nucl.\ Phys.\ B: Proc.\ Suppl., Elsevier)}
\end{center}

\vfill
\noindent
October  2010

\newpage
\thispagestyle{empty}
\vbox{}
\newpage
 
\setcounter{page}{1}


\begin{abstract}
We have just entered a new round in the testing of the flavour sector of the Standard Model 
through high-precision measurements of $B$-meson decays. A particularly exciting aspect
is the exploration of the $B_s$-meson system at LHCb. We focus on two particularly 
promising probes of new physics which may give us first solid evidence for New Physics
at the LHC: the strongly suppressed rare decay $B^0_s\to\mu^+\mu^-$ and CP-violating 
effects in the $B^0_s\to J/\psi \phi$ channel. We discuss recent theoretical developments
related to these measurements and shall also sketch other highlights of the $B$-physics
programme in the LHC era. 
\vspace{1pc}
\end{abstract}

\maketitle

\section{WHERE DO WE STAND?}
In the last decade, we have obtained many valuable new insights into flavour physics 
and CP violation through the interplay between theory and the data from the $B$ factories 
and the Tevatron. The lessons from the data collected so far is that the 
Cabibbo--Kobayashi--Maskawa (CKM) matrix is the dominant source of flavour and 
CP violation. New effects could not yet be established, although there are potential 
signals which are still not conclusive. 

The implications for the structure of New Physics (NP) is that we may actually have to deal with 
a large characteristic NP scale $\Lambda_{\rm NP}$, i.e.\ a scale that is not just $\sim\mbox{TeV}$, 
or/and that symmetries prevent large NP effects in flavour-changing neutral-current (FCNC) 
processes, where the most prominent example is ``minimal flavour violation" (MFV). It should be 
emphasized that MFV is still far from being experimentally established, and that there are various 
non-MFV scenarios with room for sizable NP effects. Prominent examples are supersymmetry, 
models with a 4th generation, warped extra dimensions, or $Z'$ models. Nevertheless, we have 
to be prepared to deal with smallish NP effects in flavour probes. 

The key problem in the use of quark-flavour physics as a probe of NP is related to the 
impact of strong interactions, leading to process-dependent, non-perturbative ``hadronic" 
parameters in the corresponding calculations. A closer look shows that the $B$-meson
system is actually a particularly promising probe: we have simplifications thanks to the
large $b$-quark mass $m_b\sim 5\,\mbox{GeV}\gg \Lambda_{\rm QCD}$, there are strategies
to determine hadronic parameters from data with the help of flavour-symmetry arguments, and 
there are tests of SM relations that could be spoiled by NP. There are two attractive ways
for NP to manifest itself in $B$ decays: contributions at the decay amplitude level to rare 
(FCNC) processes, and through contributions to $B^0_q$--$\bar B^0_q$ mixing ($q\in\{d,s\}$). 

In the following discussion, we shall focus on two topics: in Section~\ref{sec:Bsmumu}, we
have a closer look at the rare decay $B^0_s\to\mu^+\mu^-$, with a focus on a new
strategy for the branching ratio measurement proposed in Ref.~\cite{FST}, while we focus
in Section~\ref{sec:Bspsiphi} on the CP violation in $B^0_s\to J/\psi \phi$, having in particular
a critical look at the picture emerging in the Standard Model (SM) \cite{FFM}. In 
Section~\ref{sec:other}, we sketch other interesting $B$ decays, and give a brief outlook 
in Section~\ref{sec:concl}.

\section{SEARCH FOR NP IN \boldmath$B^0_s\to \mu^+\mu^-$\unboldmath}\label{sec:Bsmumu}
The decay $B^0_s\to\mu^+\mu^-$ originates from $Z$ penguins and box diagrams
in the SM, and the corresponding low-energy effective Hamiltonian takes the following
form \cite{BB}:
\begin{eqnarray}
\lefteqn{{\cal H}_{\rm eff}=-\frac{G_{\rm F}}{\sqrt{2}}\left[
\frac{\alpha}{2\pi\sin^2\Theta_{\rm W}}\right]}\nonumber\\
&&\times V_{tb}^\ast V_{ts} \eta_Y Y_0(x_t)(\bar b s)_{\rm V-A}(\bar\mu\mu)_{\rm V-A} ,
\end{eqnarray}
where $\alpha$ is the QED coupling, $\Theta_{\rm W}$ is the Weinberg angle,
$\eta_Y$ describes short-distance QCD corrections, and $Y_0(x_t\equiv m_t^2/M_W^2)$
is an ``Inami--Lim function".  Concerning the hadronic sector, only 
$\langle 0| (\bar b s)_{\rm V-A}|B^0_s\rangle$, i.e.\ the $B_s$ decay constant 
$f_{B_s}$, enters so that the $B^0_s\to\mu^+\mu^-$ channel belongs to the
cleanest rare $B$ decays. Using the data for the mass difference $\Delta M_s$ to trade 
$f_{B_s}$ into the bag parameter $\hat B_s$ yields
\begin{equation}
\frac{\mbox{BR}(B^0_s\to\mu^+\mu^-)}{\Delta M_s}=
4.4\times10^{-10}\frac{\tau_{B_s}}{\hat B_s}\frac{Y^2(\nu)}{S(\nu)},
\end{equation}
which holds in MFV models, and gives for the SM
\begin{equation}\label{BR-SM}
\mbox{BR}(B^0_s\to\mu^+\mu^-) = (3.6\pm0.4)\times 10^{-9},
\end{equation}
where the error is fully dominated by $\hat B_s$  coming from lattice QCD \cite{buras}.
As is well known, this branching ratio may be significantly enhanced through 
NP (see Ref.~\cite{buras} and references therein). 
The present 95\% C.L. upper bounds from CDF and \D0 
are still about one order of magnitude above the SM prediction in (\ref{BR-SM})  
and read as $4.3\times 10^{-8}$ \cite{CDFbound}  and 
$5.1\times 10^{-8}$ \cite{D0bound}, respectively.

The measurement of $\mbox{BR}(B^0_s\to\mu^+\mu^-)$ at LHCb will rely 
on normalization channels such as $B_u^+\to J/\psi K^+$, $B^0_d\to K^+\pi^-$ 
and/or $B_d^0 \to J/\psi K^{*0}$:
\begin{equation}
\mbox{BR}(B^0_s\to\mu^+\mu^-)
=\mbox{BR}(B_q\to X)\frac{f_q}{f_s}
\frac{\epsilon_{X}}{\epsilon_{\mu\mu}}
\frac{N_{\mu\mu}}{N_{X}}, \label{BRmumu-exp}
\end{equation}
where the $\epsilon$ are total detector efficiencies and the $N$ denote the observed 
numbers of events. The $f_q$ are fragmentation functions, which describe the probability 
that a $b$ quark fragments in a $\bar B_q$ ($q\in\{u,d,s\}$). A closer look shows that 
$f_q/f_s$ is actually the major source of uncertainty, limiting the ability to detect 
a $5 \sigma$ deviation from the SM at LHCb to 
$\mbox{BR}( B^0_s\to\mu^+\mu^-) > 11 \times 10^{-9}$ (assuming an uncertainty of
$13\%$ for $f_d/f_s$) \cite{bib:LHCbRoadMap}. 

\begin{figure}[!t]
    \centering
      \includegraphics[width=1.4in]{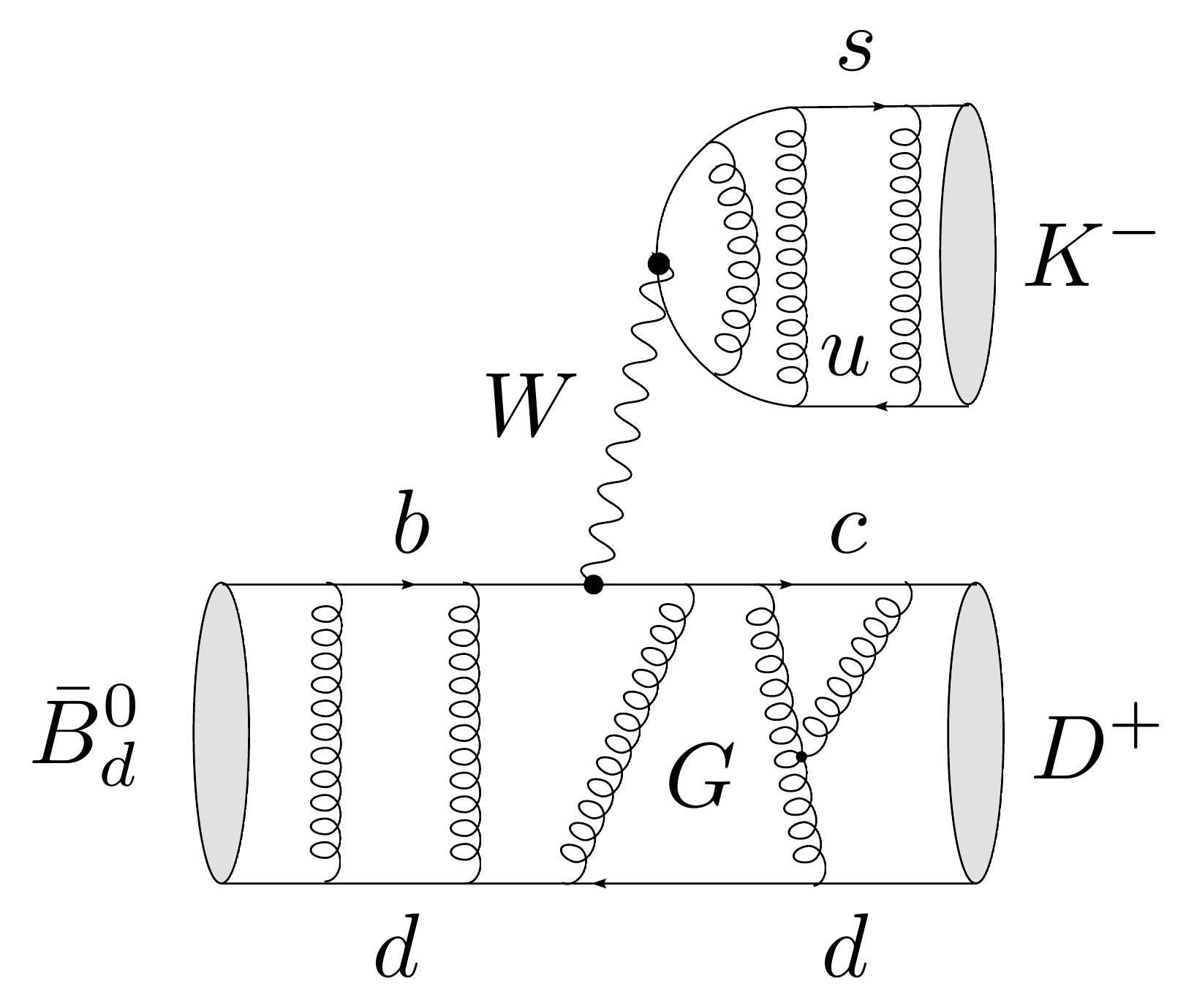} 
      \includegraphics[width=1.4in]{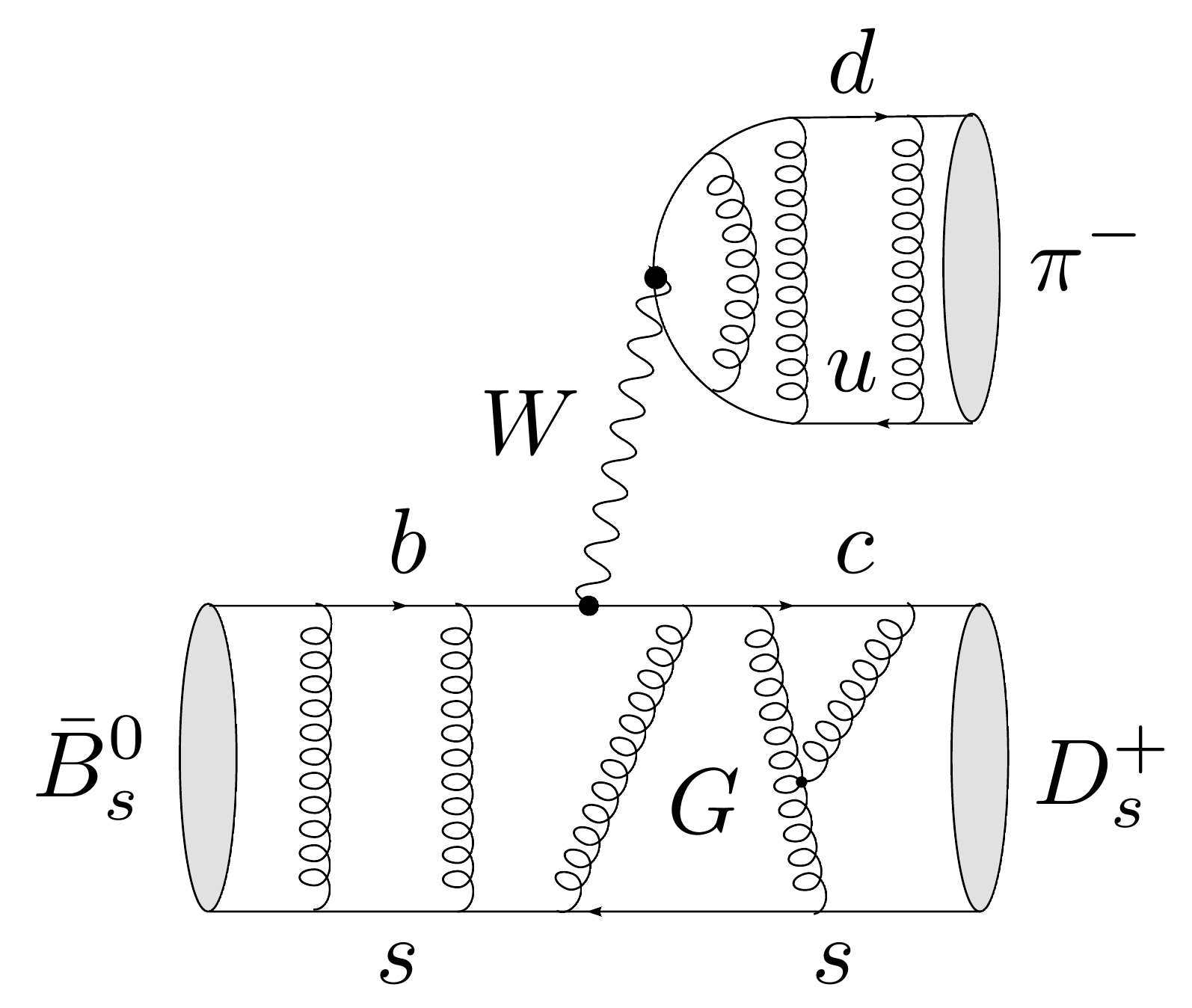}
       \caption{The topologies contributing to the $\bar B^0_d\to D^+K^-$ and 
       $\bar B^0_s\to D_s^+\pi^-$ decays.}\label{fig:diag}
 \end{figure}

Since the determinations of $f_d/f_s$ are not sufficient to meet the high precision at LHCb, 
a new strategy to measure this quantity at LHCb was proposed in Ref.~\cite{FST}. 
The starting point is 
\begin{equation}\label{eq:simple}
\frac{N_s}{N_d} = \frac{f_s}{f_d}\times \frac{\epsilon(B_s\rightarrow
  X_{1})}{\epsilon(B_d\rightarrow X_{2})}\times \frac{\mbox{BR}(B_s\rightarrow 
  X_{1})}{\mbox{BR}(B_d\rightarrow X_{2})};
\end{equation}
knowing the ratio of the branching ratios, we could obviously extract $f_d/f_s$ 
experimentally. In order to implement this feature in practice, the $B_s\to X_1$ and $B_d\to X_2$
decays have to satisfy the following requirements:
\begin{itemize}
\item the ratio of their branching ratios must be easy to measure at LHCb;
\item the decays must be robust with respect to the impact of NP contributions;
\item the ratio of their BRs must be theoretically well understood  within the SM.
\end{itemize}
These requirements guide us to the $\bar B_s^0\rightarrow D_s^+ \pi^-$ and 
$\bar B_d^0 \rightarrow D^+ K ^-$ channels, which receive only contributions from 
colour-allowed tree-diagram-like topologies, as can be seen in Fig.~\ref{fig:diag}. 
Their hadronic amplitudes are related to
each other by the $U$-spin symmetry of strong interactions, and the decays are known
as prime examples of decays where ``factorization" is expected to hold:
\begin{eqnarray}
\lefteqn{{\cal A}(\bar B^0_q \to D_q^+P^-)
=\frac{G_{\rm F}}{\sqrt{2}} V_{q}^\ast V_{cb}}\nonumber\\
&&\times a_1(D_qP) f_P F_0^{(q)}(m_P^2)(m_{B_q}^2-m_{D_q}^2).
\end{eqnarray}
This feature could be put on a rigorous theoretical basis in the heavy-quark 
limit  \cite{BBNS,SCET}. In QCD factorization, $a_1$ is found as a quasi-universal 
quantity $|a_1|\simeq 1.05$ with very small process-dependent  ``non-factorizable" 
corrections \cite{BBNS}. 

So far, this interesting feature did not have any practical
application. However, we can actually use these decays for the determination
of $f_d/f_s$ at LHCb. On the one hand, we have 
\begin{eqnarray}
\lefteqn{\frac{\mbox{BR}(\bar B^0_s\to  D_s^+\pi^-)}{\mbox{BR}
(\bar B^0_d\to D^+K^-)}\sim \frac{\tau_{B_s}}{\tau_{B_d}}
\left|\frac{V_{ud}}{V_{us}}\right|^2 }\nonumber\\
&&\times
\left(\frac{f_\pi}{f_K}\right)^2\left[\frac{F_0^{(s)}(m_\pi^2)}{F_0^{(d)}(m_K^2)}
\right]^2\left|\frac{a_1(D_s\pi)}{a_1(D_dK)}\right|^2,\label{eq-rat}
\end{eqnarray}
while the ratio of the number of signal events observed in the experiment is given by
\begin{equation}
\frac{N_{D_s\pi}}{N_{D_d K}}=\frac{f_s}{f_d} 
\frac{\epsilon_{D_s\pi}}{\epsilon_{D_d K}}
\frac{\mbox{BR}(\bar B^0_s \to D_s^+\pi^-)}{\mbox{BR}(\bar B^0_d \to D^+K^-)}.
\end{equation}
Consequently, we obtain
\begin{equation}\label{fs-det}
\frac{f_d}{f_s}=12.88\times\frac{\tau_{B_s}}{\tau_{B_d}}\times
\left[{\cal N}_a {\cal N}_F
\frac{\epsilon_{D_s \pi}}{\epsilon_{D_d K}}
\frac{N_{D_d K}} {N_{D_s\pi}}\right],
\end{equation} 
with
\begin{equation}\label{NF_definition}
 {\cal N}_a \equiv \left|\frac{a_1(D_s\pi)} {a_1(D_dK)}\right|^2,
 \quad {\cal N}_F \equiv \left[\frac{F_0^{(s)}(m_\pi^2)}{F_0^{(d)}(m_K^2)}\right]^2.
\end{equation}

The $\bar B^0_s\to D_s^+\pi^-$ and $\bar B^0_d\to D^+K^-$ decays can be 
exclusively reconstructed with the help of the $D^+\to K^-\pi^+\pi^+$ and $D_s^+\to K^+K^-\pi^+$ 
transitions, respectively. Since both channels are selected with an identical flavour final 
state containing the four charged hadrons $KK\pi\pi$, the uncertainty on
$\epsilon_{D_s \pi}/\epsilon_{D_d K}$ is small. Using a toy Monte Carlo simulation to generate
a $0.2$~fb$^{-1}$ sample  yields about 5500 
 $\bar B^0_s\to D_s^+\pi^-$ and 1100 $\bar B^0_d\to D^+K^-$ events, resulting in
 an error of $7.5\%$ for $r\equiv (\epsilon_{D_s \pi} N_{D_d K})/(\epsilon_{D_d K} N_{D_s\pi})$.
 Here the dominant uncertainty comes from $\mbox{BR}(D_s\to K^+K^-\pi)= (5.50 \pm 0.28)\%$.
Extrapolating to  $1$~fb$^{-1}$, which corresponds to the end of 2011, the statistical
uncertainty becomes essentially negligible so that the total uncertainty is reduced to 
$\Delta r \sim5.6\%$.

Concerning the theoretical uncertainties, we have to deal with non-factorizable
$U$-spin-breaking effects, which are described by 
\begin{equation}
{\cal N}_a  \approx 1+ 2\Re(a_1^{\rm NF}(D_s\pi)-a_1^{\rm NF}(D_dK)).
\end{equation}
Here $a_1^{\rm NF}$ is associated with non-universal, i.e.\ process-dependent,
non-factorizable contributions, which cannot be calculated reliably. However, they 
arise as power corrections to the heavy-quark limit, i.e.\ they are suppressed by at 
least one power of $\Lambda_{\rm QCD}/m_b$,  and are -- in the decays at hand -- 
numerically expected at the few percent level \cite{BBNS}. Moreover, since we are
only sensitive to an $SU(3)$-breaking difference, $1-{\cal N}_a$ is conservatively 
expected to be at most a few percent. In this context, it is important to emphasize that
we can also experimentally test factorization, as discussed in detail in Ref.~\cite{FST}.

The major uncertainty affecting (\ref{fs-det}) is hence the 
form-factor ratio ${\cal N}_F$, where $U$-spin-breaking corrections 
arise from $d$ and $s$ spectator-quark effects. Unfortunately, the 
$B_s\to D_s$ form factors have so far received only small theoretical attention. 
In Ref.~\cite{chir}, such effects were explored using heavy-meson chiral 
perturbation theory, while QCD sum-rule techniques were applied in 
Ref.~\cite{BCN}. The numerical value given in the latter paper yields 
${\cal N}_F=1.3\pm0.1$. If we assume ${\cal N}_{F}>1$ (as the radius of the $B^0_s$ 
is smaller than that of the $B^0_d$), we obtain the following lower bound
\begin{equation}\label{BR-bound}
\mbox{BR}(B^0_s\to\mu^+\mu^-)>\underbrace{\mbox{BR}(B^0_s\to\mu^+\mu^-)_0}_{\mbox{assumes ${\cal N}_{F}=1$}},
\end{equation}
which offers an interesting probe for NP. In order to match experiment, it is sufficient to calculate the $U$-spin-breaking corrections to $F_0^{(s)}(m_\pi^2)/F_0^{(d)}(m_K^2)$ with non-perturbative
methods, such as lattice QCD, at the level of 20\%, which should be feasible soon.

\begin{figure}
  \centering
  \begin{tabular}{cc}
    \includegraphics[width=0.240\textwidth]{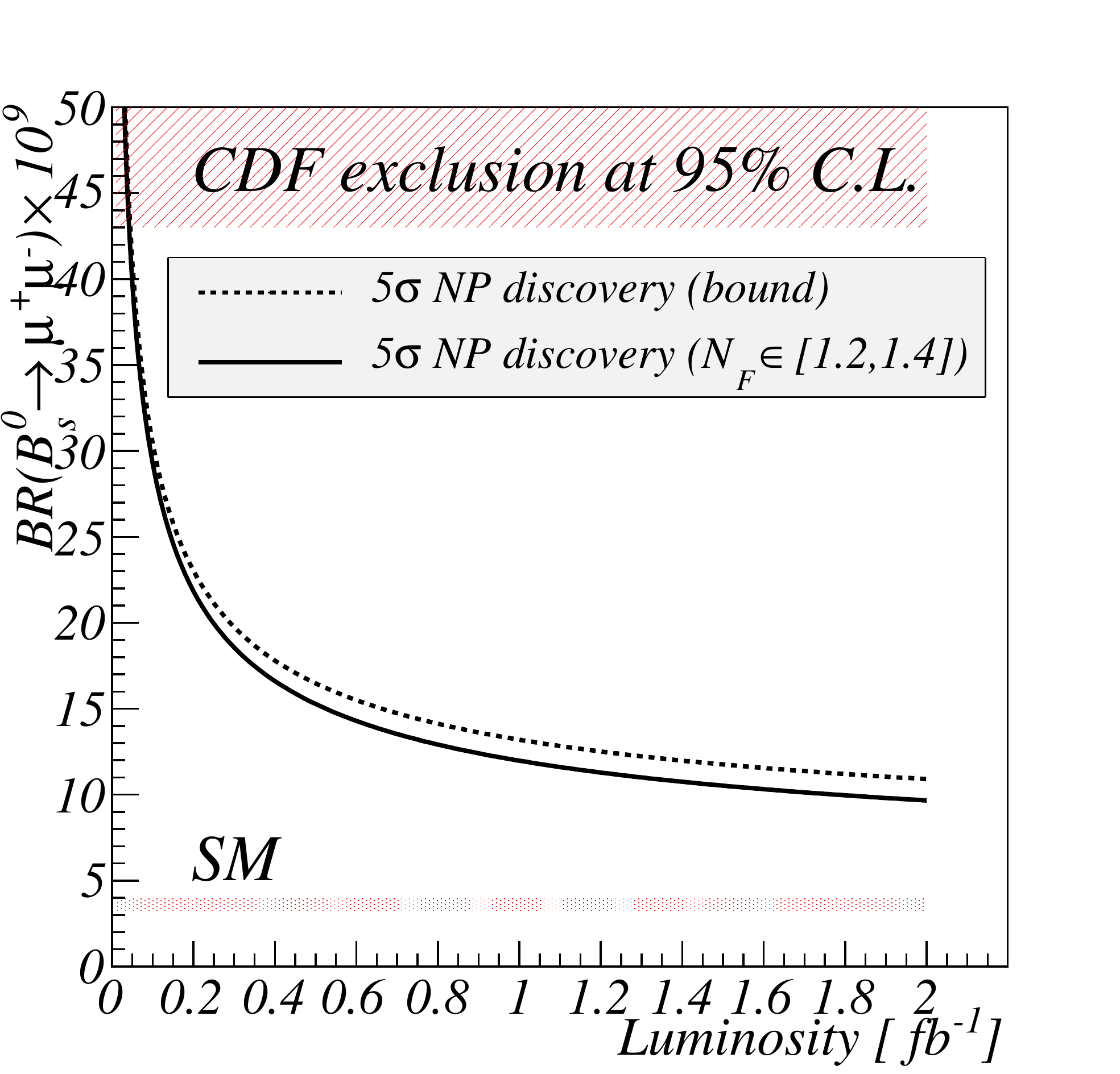} 
    \includegraphics[width=0.240\textwidth]{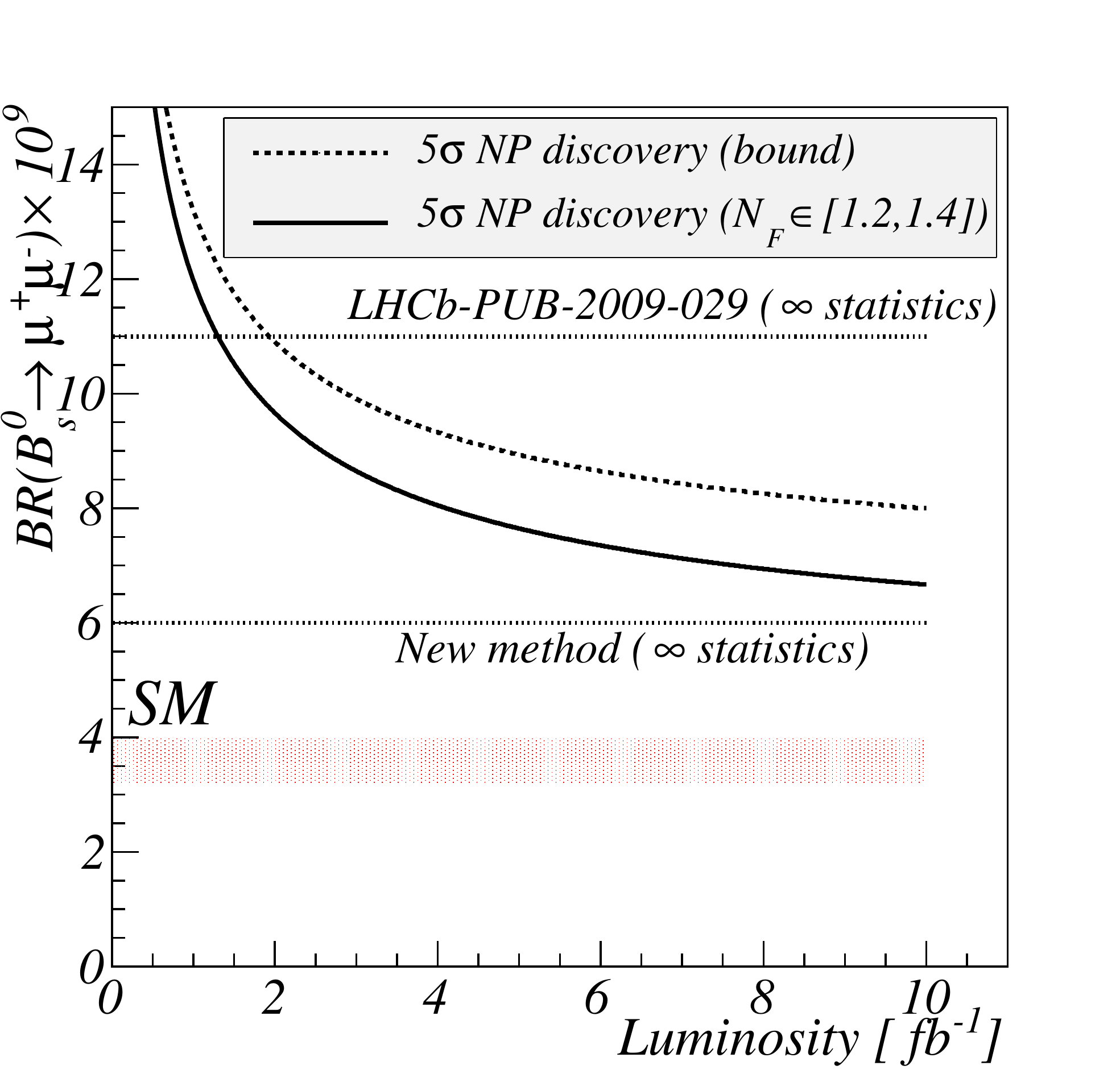} 
  \end{tabular} 
  \caption{Illustration of the NP discovery potential in  $B^0_s\to\mu^+\mu^-$
   at LHCb resulting from the strategy proposed in \cite{FST}, as discussed 
  in the text.}\label{fig:NP}
\end{figure}

In Fig.~\ref{fig:NP}, we illustrate the NP discovery potential of the $B^0_s\to\mu^+\mu^-$ channel 
at LHCb resulting from the strategy proposed in Ref.~\cite{FST}. Here we show the smallest 
value of $\mbox{BR}(B^0_s\to\mu^+\mu^-)$ that allows the detection of a $5 \sigma$ deviation 
from the SM as a function of the luminosity  at LHCb (at the nominal beam  energy of 14 TeV). 
The figure on the left-hand side shows the low-luminosity regime, whereas the one on the 
right-hand side illustrates the asymptotic behaviour. The plot on the right-hand shows that 
we obtain a NP discovery potential about twice as large as the present LHCb 
expectation~\cite{bib:LHCbRoadMap} (upper horizontal line) enabling a possible
discovery of NP down to $\mbox{BR}(B^0_s\to\mu^+\mu^-) > 6\times 10^{-9}$ 
(lower horizontal line).
In addition to the increased sensitivity in the regime of low branching ratios, 
even for large values close to the current CDF exclusion limit the significance 
of a possible NP discovery would be increased. Thanks to the decrease of the 
systematical uncertainty, LHCb will be able to fully exploit the statistical 
improvement, taking full advantage of the accumulated LHCb 
data up to 10~fb$^{-1}$, which corresponds to five years of nominal LHCb data taking.
The value of $f_d/f_s$ is not only crucial for the measurement of 
$\mbox{BR}(B^0_s\to \mu^+\mu^-)$ but enters the measurement of any $B_s$ branching
ratio at LHCb.

\section{SEARCH FOR NP IN \boldmath$B^0_s\to J/\psi\phi$\unboldmath}\label{sec:Bspsiphi}
The $B^0_s\to J/\psi \phi$ channel is the $B_s$-meson counterpart of the 
$B^0_d\to J/\psi K_{\rm S}$ decay. NP may well manifest itself  in CP-violating phenomena 
in $B^0_s\to J/\psi \phi$ through contributions to $B_s^0$--$\bar B_s^0$ mixing, yielding a 
mixing phase $\phi_s$ different from the doubly Cabibbo-suppressed SM value
$\phi_s^{\rm SM}\approx-2^\circ$; since the final state of $B^0_s\to J/\psi \phi$ is an 
admixture of CP-even and CP-odd eigenstates, a time-dependent angular analysis 
is required to search for NP effects \cite{DDFN}. The most recent compilation of the 
corresponding results by the CDF and \D0 collaborations can be found in 
Refs.~\cite{CDF-phis} and \cite{D0-phis}, respectively. Unfortunately, the 
situation is not conclusive, though the CDF and \D0 analyses are consistent with
each other. While CDF  finds 
$\phi_s\in [-59.6^\circ,-2.29^\circ]\sim-30^\circ \, \lor \, [-177.6^\circ,-123.8^\circ]\sim-150^\circ$
(68\% C.L.), \D0  gives a best fit value around $\phi_s\sim -45^\circ$, taking also information 
from the dimuon charge asymmetry and the measured $B_s^0\to D_s^{(*)+}D_s^{(*)-}$ 
branching ratio into account.

The experimental prospects for the analysis of $B^0_s\to J/\psi \phi$ at LHCb are
very promising. With $2$~fb$^{-1}$, an experimental uncertainty of 
$\sigma(\phi_s)_{\rm exp}\sim 1^\circ$ can be achieved, which could be reduced at an
LHCb upgrade with an integrated luminosity of 100\,$\mbox{fb}^{-1}$ to 
$\sigma(\phi_s)_{\rm exp}\sim 0.2^\circ$.

\begin{figure}
   \centering
   \includegraphics[width=6.5truecm]{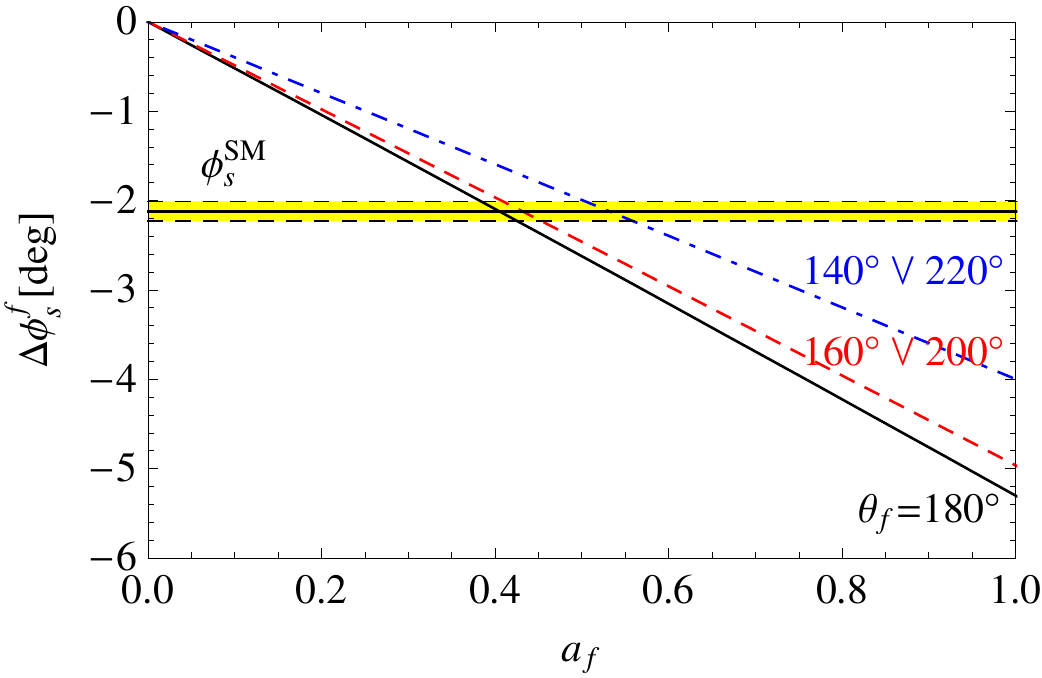} 
    \caption{Dependence of the hadronic phase shift $\Delta\phi^f_s$ on $a_f$ for various 
   values of $\theta_f$ \cite{FFJM}.}\label{fig:DelPhi}
\end{figure}

So far, the SM penguin effects were fully neglected in the analysis of the CP violation
in the $B^0_s\to J/\psi \phi$ channel:
\begin{equation}
\xi^{(s)}_{(\psi\phi)_f}\,\propto\, e^{-i\phi_s}
\bigl[1-2\,i\,\lambda^2 a_f e^{i\theta_f}\sin\gamma + {\cal O}(\lambda^4)\bigr],
\end{equation}
where $a_f e^{i\theta_f}$ describes the ratio of penguin to tree contributions for a given
final-state configuration $(J/\psi\phi)_f$. In Ref.~\cite{FFM}, a detailed discussion of
their impact was given, proposing also a strategy to control these penguin effects
through data. 
The penguin contributions modify the expression for the mixing-induced CP violation
$\hat A_{\rm M}^f$ as follows:
\begin{equation}\label{AM}
{\eta_f \hat A_{\rm M}^f}/{\sqrt{1-(\hat A_{\rm D}^f)^2}}=
\sin(\phi_s+\Delta\phi^f_s),
\end{equation}
where $\eta_f$ is the CP eigenvalue of the final-state configuration, $\hat A_{\rm D}^f$
is a direct CP asymmetry (which can be measured), and $\Delta\phi^f_s$ is a hadronic
phase shift caused by the penguin contributions, which can be expressed in terms
of $a_f$ and $\theta_f$ as given in Ref.~\cite{FFM}. It should be stressed that 
$\Delta\phi_s^f$ does not depend on the value of $\phi_s$ itself. In Fig.~\ref{fig:DelPhi}, 
we show the resulting dependence of $\Delta\phi_s^f$ on the penguin parameter $a_f$ 
for various values of $\theta_f$. We observe that $\Delta\phi^f_s$ is of the same
size as $\phi_s^{\rm SM}$ for $a_f\sim0.4$. As far as the direct CP asymmetry 
is concerned, we have $-0.05 \lsim \hat A_{\rm D}^f \lsim +0.05 $ for 
$a_f\lsim1$ and values of $|\theta_f-180^\circ|$ as large as $40^\circ$ \cite{FFM}.
As we expect $\cos\theta_f<0$, the shift of $\phi_s$ is expected 
to be negative as well, i.e.\ it would interfere constructively with $\phi_s^{\rm SM}$.
These features are fully supported by a recent analysis of the $B^0_d\to J/\psi \pi^0$
channel \cite{FFJM} (see also Ref.~\cite{CPS}).

Consequently, it is important to get a handle on the penguin effects in the 
$B^0_s\to J/\psi\phi$ decay. This can be done by means of the 
$B^0_s\to J/\psi \bar K^{*0}$ mode, which has the following SM amplitude:
\begin{equation}
A(B_s^0\to (J/\psi \bar K^{*0})_f)\propto 1-a_f' e^{i\theta_f'} e^{i\gamma}.
\end{equation}
It should be stressed that the penguin term is here {\it not} doubly Cabibbo-suppressed. 
If we use the $SU(3)$ flavour symmetry and neglect 
penguin annihilation and exchange amplitudes (which can be probed through
$B^0_d\to J/\psi \phi$), we have $a_f=a_f'$ and $\theta_f=\theta_f'$. In the summer of 2010,
CDF has announced the observation of the $B^0_s\to J/\psi \bar K^{*0}$ mode, with a 
branching ratio at the $8\times 10^{-5}$ level \cite{CDF-obs}. Moreover, also the
$B^0_s\to J/\psi K_{\rm S}$ decay was observed, which allows us to control the penguin 
effects discussed above in the measurement of $\sin2\beta$ through $B^0_d\to J/\psi K_{\rm S}$
\cite{RF-BspsiK,BFK}.

The determination of the penguin parameters from the observables of the angular
distribution of $B^0_s\to J/\psi \bar K^{*0}$ is presented in Ref.~\cite{FFM}. Let us here
just emphasize that the favoured negative sign of $\Delta\phi_s$ implies a constructive
interference with $\phi_s^{\rm SM}\sim-2^\circ$ in (\ref{AM}). For values of
$a_f'=0.4$ and $\theta_f'=220^\circ$ (consistent with the picture following from 
current $B^0_d\to J/\psi \pi^0$ data \cite{FFJM}), we get a phase shift of 
$\Delta \phi_s^f=-1.7^\circ$, which yields $\eta_f\hat A_{\rm M}^f=-6.7\%$, i.e.\  about 
twice the naive SM value. Consequently, without a control of the penguin effects, 
this SM effect would be misinterpreted as a $4\,\sigma$ NP effect with $2\,\mbox{fb}^{-1}$ 
at LHCb, and about  $20\,\sigma$ at an upgrade with $100\,\mbox{fb}^{-1}$. Should 
we find large mixing-induced CP violation in $B^0_s\to J/\psi \phi$, such as
$\eta_f\hat A_{\rm M}^f\sim-40\%$, we would have an immediate and unambiguous 
signal of NP. On the other hand, should we find $\eta_f\hat A_{\rm M}^f\sim-(5...10)\%$, 
more theoretical and experimental work would be needed in order to settle the picture.
Also in the case of $B^0_d\to J/\psi K_{\rm S}$, we have to control the penguin effects
in order to match the experimental precision at LHCb \cite{BFK}.

\section{OTHER \boldmath$B$ PHYSICS TOPICS\unboldmath}\label{sec:other}
There is much more interesting physics left for the $B$-decay studies at LHCb. 
An important line of research is given by precision measurements of the angle 
$\gamma$ of the unitarity triangle. This quantity can be determined through pure 
``tree" decays on the one hand (such as $B^0_s\to D_s^\mp K^\pm$), and through 
decays with penguin contributions on the other hand 
($B^0_s\to K^+K^-$, $B^0_d\to\pi^+\pi^-$ system). The central question is whether 
we will get values of $\gamma$ that are consistent with one another \cite{vagnoni}.

Another key topic is given by the study of rare decays, complementing the leptonic
$B^0_s\to\mu^+\mu^-$ (and its even stronger suppressed partner $B^0_d\to\mu^+\mu^-$)
channel discussed above. The semileptonic decays 
$B^0_d\to K^{*0}\mu^+\mu^-$, $B^0_s\to \phi \mu^+\mu^-$ offer another interesting
probe for NP. Here the hadronic sector involves quark-current form factors, and the 
goal is to find and measure observables that are particularly robust with respect to 
the corresponding uncertainties. The prime example is the 0-crossing of the
forward--backward asymmetry; other observables were recently proposed (see Ref.~\cite{egede}).
There are also non-leptonic rare decays that originate only from loop processes. 
Particularly interesting are CP-violating asymmetries in $B^0_s\to\phi\phi$ and similar 
modes. In order to deal with hadronic corrections, flavour symmetries 
offer strategies to control them through experimental data. Also here the key question is
whether we will encounter discrepancies with respect to the SM picture. 

Studies of charm physics offer another line of research. While FCNCs in the $B$ 
system are sensitive to new effects in the up sector, charm physics probes the down 
sector, i.e.\ we have $b$, $s$, $d$ quarks running in the SM loops. Such a
process is $D^0$--$\bar D^0$ mixing, which is seen in the ball park of the 
SM. NP could be hiding there, but is obscured through long-distance QCD effects. In 
order to search for NP, CP-violating effects, which are tiny in the SM but may be enhanced 
through NP contributions, are most promising.

Last but not least, we can also search for lepton flavour violation through 
$B^0_{d,s}\to e^\pm \mu^\mp$ and  $B^0_{d,s}\to \mu^\pm\tau^\mp$ decays, which are
forbidden in the SM but may arise in NP scenarios. These studies complement 
other searches by means of $\mu\to e \gamma$,
$\tau\to\mu\gamma$ or $\tau\to \mu\mu\mu$ processes.

\section{OUTLOOK}\label{sec:concl}
We are currently moving towards new frontiers in precision $B$ physics thanks to the
start-up of the LHCb experiment. The last decade has led to various interesting
results, showing -- among many other insights -- that the CKM matrix is the dominant
source of flavour and CP violation. Potential signals of new phenomena were also
seen, although the situation is still not conclusive. 

Flavour physics takes part in the big scientific adventure of this decade, which is the LHC. 
Specific NP scenarios still leave room for sizable effects. Particularly promising channels to 
find first signals at LHCb (and the LHC) are $B^0_s\to \mu^+\mu^-$ and $B^0_s\to J/\psi \phi$. 

In view of the new territory we are about to enter now, the SM 
phenomena have to be critically reviewed and strategies to control the corresponding 
hadronic uncertainties to be further developed and refined. Concerning the
search for NP, the patterns in specific scenarios should be further explored. In particular
correlations between different observables should play a key role in revealing the structure
of NP, should we actually see footprints of physics beyond the SM. 
Moreover, synergies with the high-$Q^2$ physics at ATLAS and CMS should be further 
studied and exploited. Exciting times are ahead of us!

\noindent
{\it Acknowledgements}\\
I would like to thank the organizers, in particular Giulia Ricciardi, for hosting this 
exciting conference.  I am also grateful to Nicola Serra and Niels Tuning for 
comments on the manuscript and a most enjoyable collaboration.

\end{document}